\documentclass[aps,pre,amsmath,amsfonts,nofootinbib]{revtex4-2}
\usepackage{xcolor}
\usepackage{graphicx}
\usepackage[mathlines, switch]{lineno}
\begin{document}
	


\title[DCAlign v1.0]{DCAlign v1.0: Aligning biological sequences using co-evolution models and informed priors}
\author{Anna Paola Muntoni\,$^{\text{1,2}*}$, Andrea Pagnani\,$^{\text{1,2,3}}$}
\address{$^{\text{\sf 1}}$Italian Institute for Genomic Medicine, IRCCS Candiolo, SP-142, I-10060 Candiolo (TO), Italy \\
$^{\text{\sf 2}}$Politecnico di Torino, Corso Duca degli Abruzzi, 24, I-10129, Torino, Italy \\
$^{\text{\sf 3}}$INFN, Sezione di Torino, Torino, Italy 
}
\begin{abstract}\textbf{Summary:} DCAlign is a new alignment method able to cope with the conservation and the co-evolution signals that characterize the columns of multiple sequence alignments of homologous sequences. However, the pre-processing steps required to align a candidate sequence are computationally demanding. We show in v1.0 how to dramatically reduce the overall computing time by including an empirical prior over an informative set of variables mirroring the presence of insertions and deletions.\\
\textbf{Availability and implementation:} DCAlign v1.0 is implemented in Julia and it is fully available at \url{https://github.com/infernet-h2020/DCAlign} \\
\textbf{Contact:} anna.muntoni@polito.it \\ 
\textbf{Supplementary information:} Supplementary data are available at \textit{Bioinformatics}
online.
\end{abstract}

\maketitle

\section{Introduction}

A common task in Bioinformatics is to cast evolutionary-related
biological sequences into a Multiple Sequence Alignment (MSA). The
objective of this task is to identify and align conserved regions of
the sequences by maximizing the similarity among the columns of the
MSA.  State-of-the-art alignment methods, like {\tt HMMER} for proteins
\citep{eddy2011}, and {\tt Infernal} \citep{nawrocki2013} for RNAs, use hand-curated MSAs of small representative subsets of  sequences to be aligned (the so-called
\textit{seed} alignments). Whereas for proteins, {\tt HMMER} builds the Hidden Markov Model (HMM) by using only the seed alignment, {\tt Infernal} needs also secondary structure information to generate a Covariance Model (CM).  In both cases, HMM (for proteins) or CM (for RNAs) are used to align query sequences. 
However, homologous sequences show signals of correlated mutations (epistasis) undetected by profile models. \\
Conservation and co-evolution signals are at the basis of Direct
Coupling Analysis (DCA)-based statistical models \citep{morcos2011, cocco2018}. 
Recently, these models have been used to align biological sequences \citep{muntoni2020} and perform remote homology search \citep{wilburn2020} by alignment of the sequences to a seed model, or by pairwise
alignments of seed models \citep{talibart2021}.
The method in \citep{muntoni2020}, viz. {\tt DCAlign}, returns
the ordered sub-sequence of a query unaligned sequence which maximizes an objective function related to the DCA model of the seed.
In this latter case, standard DCA models fail to adequately describe the statistics of insertions and gaps.  \\
To alleviate this limitation, we added to the objective function
gap and insertion penalties learned from the seed alignment. While
for the insertions, the computational complexity is negligible,
inferring gap penalties is a time-consuming problem (see
\citep{muntoni2020} and \textcolor{blue}{Supplementary text}). Here,
we treat penalties in terms of informed priors computed
from the seed sequences.
The parameters for gaps and insertions, extracted from the seed alignment, are determined in an unsupervised manner. Finally, to further
speed up the learning of the seed-based objective function, we obtain
the parameters of the DCA model using pseudo-likelihood maximization
\citep{ekeberg2013} instead of Boltzmann Machine Learning
\citep{figliuzzi2018, muntoni2021}. {\tt DCAlign v1.0}, is a computational
pipeline that allows for the computation of the seed-model parameters in a few minutes, contrary to its original implementation which required at
least a day of computation in the best scenario. The alignment problem is then solved approximately through a message-passing
algorithm (see \textcolor{blue}{Supplementary text}).


\section{Methods}

\begin{figure*}[h]
\centering
\includegraphics[width=\textwidth]{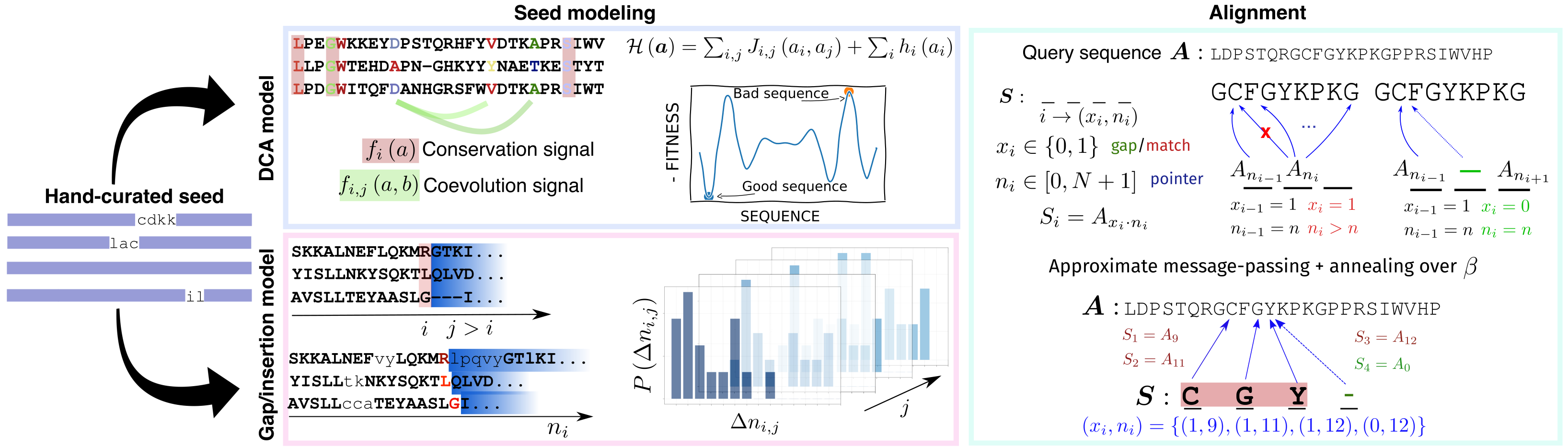}
\caption{Schematic representation of the {\tt DCAlign v1.0} pipeline. From a (given) hand-curated alignment (the seed, shown in the left panel), our algorithm learns (i) a DCA model $\mathcal{H}$ exploiting the one-site and two-site statistics of the seed (upper central box), and (ii) the gap and insertion penalties by means of the empirical distribution of the pointer differences $P\left(\Delta n_{ij}\right)$ for $i=1, \ldots, L$, and $j>i$ (bottom central box). The three sequences represent the three scenarios that can occur between position $i$ and $j = i+1$: some insertion can appear, no insertion and no gap is present, or $i+1$ contain a gap, so $\Delta n_{i,i+1} = 0$. For $j > i+1$ (lighter blue cases), both insertions and matched symbols contribute to the computation of the $\Delta n_{i,j}$, while gaps do not carry any contribution (see \textcolor{blue}{Fig. S1} for a more detailed example). 
The alignment problem is then mapped into a constrained optimization problem over the $\left(\boldsymbol{x},\boldsymbol{n}\right)$ variables. The constraints on the variables and an example of alignment are shown in the right panel. \label{fig:one}}
\end{figure*}
Our alignment algorithm estimates the optimal ordered sub-sequence compatible with a DCA model and empirical knowledge of insertions and gaps of the seed. Let $\boldsymbol{A}$ be an unaligned sequence of length $N$, and $\boldsymbol{S}$ be its aligned counterpart of length $L$ (which is the length of the seed MSA). We only consider the $L \le N$ case. At each $i = 1, \ldots, L$, we define a Boolean variable $x_{i} \in \{ 0,1 \}$ and a pointer $n_{i} \in \{ 0, \ldots , N+1 \}$. The variable $x_{i}$ indicates whether the position $i$ is a \textit{gap} `-` ($x_{i} = 0$) or a \textit{match}, {\em i.e.} a symbol in $\boldsymbol{A}$. When $i$ is a match, $n_{i}$ identifies where $S_{i}$ matches $\boldsymbol{A}$, i.e. $S_{i} = A_{n_{i}}$; instead, for $x_{i} = 0$, the value of $n_{i}$ is used for keeping track of the last matched symbol in $\boldsymbol{A}$. 
Let us define a pointer-difference variable as $\Delta n_{i,j} =
n_{j}-n_{i}$ for $i = 1,\ldots,L$ and $j>i$. Each auxiliary variable
$\Delta n_{i,j}$ quantifies how many symbols of the unaligned sequence
$\boldsymbol{A}$ are present between two $i,j$ positions of the
aligned counterpart $\boldsymbol{S}$. If a configuration of the $\boldsymbol{n}$ is given, the full set of the pointer differences reveal the presence of insertions and gaps between any columns $i$ and $j$ of the alignment (see \textcolor{blue}{Supplementary text}).
\subsection{Seed modeling}
Together with a DCA model of the aligned seed (see Fig.\ref{fig:one}, central panel), for every site $i$ (in red), we compute the $\Delta n_{i,j}$ for $j > i$ for all the seed sequences, and we learn an empirical probability $P_{i,j}\left( \Delta n_{i,j}\right)$ as shown in the bottom central panel of Fig.\ref{fig:one} (this procedure is computationally very fast). The color gradient is associated with the value of $j$, the lighter the color, the larger is $j$. In Fig.\ref{fig:one} (bottom central panel) we consider as an example three sequences differing in the nature of the $\Delta n_{i,i+1}$. 
\subsection{Alignment procedure}
We can express the alignment problem in terms of the following optimization problem:
\begin{equation} 
	\boldsymbol{x}, \boldsymbol{n} =
\mathrm{argmax}_{\boldsymbol{\bar{x}}, \boldsymbol{\bar{n}}}
\frac{e^{-\beta \mathcal{H}\left(\boldsymbol{\bar{x}},
  \boldsymbol{\bar{n}}\right)}}{Z(\beta)}
\prod_{i,j}P_{ij}^{\beta}\left(\boldsymbol{\bar{n}}\right),
\end{equation}
where $\mathcal{H}$ is the DCA model describing the seed (see Fig. \ref{fig:one}, top central panel), $Z$ is a normalization factor, and $\beta$ is a free parameter whose relevance will be discussed below. The maximization only runs over the feasible assignment of the variables, i.e. we impose that $n_{i+1} > n_{i}$ for every column $i$. The informed prior will guide the optimization process towards solutions that, among those that maximize the Boltzmann distribution associated with $\mathcal{H}$, reproduce the statistics of the seed pointer differences. Unfortunately, the problem thus stated is unfeasible as the normalization function $Z$ cannot be efficiently computed. Similarly to the first {\tt DCAlign} version, we
use an approximate message-passing algorithm coupled with an annealing scheme over $\beta$ (i.e. we iteratively increase $\beta$) to get the best alignment for the query sequence $\boldsymbol{A}$ (see \textcolor{blue}{Supplementary text} and \textcolor{blue}{Fig. S2}).


\section{Results}

We can classify the type of tests performed to assess the performance of our computational strategy into three different categories:
\begin{itemize}
\item {\em Comparison with the previous implementation:} As in \citep{muntoni2020}, we compared our results against {\tt HMMER},  {\tt Infernal} (the last algorithm only for RNA sequences) on four Pfam (PF00035, PF00677, PF00684, PF00763), and Rfam (RF00059, RF00162, RF00167, RF01734) families. A detailed description of the  dataset is contained in \textcolor{blue}{Tabs. S2-3}. We utilized the following comparison metrics: (i) the positive predictive value (PPV) of the DCA-based contact prediction \citep{morcos2011,cocco2018}, (ii) the proximity measures between the generated and the seed MSAs. As far as the contact map prediction is concerned, we observe either a mild improvement or a similar performance. With respect to the proximity measures, we notice a negligible increase in the average distance between seed sequences and generated alignments (see \textcolor{blue}{Figs. S3-6}, and \textcolor{blue}{Tabs. S7-10}).
\item {\em Leave-one-out experiment:} As a stress test for {\tt DCAlign v1.0} we also compared our results to twenty-five ground-truth MSAs either extracted from benchmark sets \citep{bahr2001, thompson2005, freyhult2007} or built from structural alignments \citep{akdel2020} (see \textcolor{blue}{Tabs. S2}, \textcolor{blue}{S4-5}). The numerical experiments consist of iteratively excluding one of the sequences of the reference alignment and training HMM, CM, or DCAlign using the remaining sequences. The excluded sequence is then aligned and quantitatively compared to the ground truth (viz. the structural alignment, or the benchmark sets). The emerging picture depends on the data type considered: for benchmark sets all computational strategies seem to perform reasonably well. In particular, HMMER (resp. Infernal) and our algorithm provide similar outcomes for protein (resp. RNA) domains (see \textcolor{blue}{Figs. S7-10}, \textcolor{blue}{Tabs. S11-12}). However, when we consider structural alignments as our reference ground truth, our method significantly outperforms HMMER as shown in \textcolor{blue}{Figs. S11-12} and \textcolor{blue}{Tabs. S13-14}. 
\item{\em Divergent sequence alignment:} Finally, to assess our algorithm's remote homology detection performance, we considered three RNA benchmark sets (the seed of Rfam RF00162 \citep{kalvari2020}, Twister type P1 \citep{roth2014}, tRNA \citep{sprinzl1998}, see \textcolor{blue}{Tab. S6}) from \citep{wilburn2020}. Results suggest that Infernal is the best-performing method on two of the three datasets, while our method achieves the best alignment for the tRNA case. Note that Infernal is trained using secondary structure information that our algorithm does not use. All results are presented in \textcolor{blue}{Fig. S13} and \textcolor{blue}{Tab. S15}.
\end{itemize}

From a computational efficiency point of view, the time needed to train the algorithm is significantly smaller than both our old implementation and CM-Infernal (see \textcolor{blue}{Supplementary text} and  \textcolor{blue}{Fig. S14}). However, the time necessary to align a sequence is equivalent compared to {\tt DCAlign}, and probably to other computational strategies taking into account epistasis \citep{wilburn2020,talibart2021}. 
\section{Conclusion}
{\tt DCAlign v1.0} is a new implementation of the DCA-based alignment
technique, {\tt DCAlign}, which conversely to the first implementation,
allows for a fast parametrization of the seed alignment. The new
modeling significantly drops the pre-processing time and guarantees a
qualitatively equivalent alignment of a set of target sequences.
\section*{Acknowledgements}
APM and AP acknowledge financial support from Marie
Sk{\l}odowska-Curie, grant agreement no. 734439(INFERNET). We also
warmly thank Indaco Biazzo, Alfredo Braunstein, Louise Budzynski and
Luca Dall'Asta for interesting discussions.\vspace*{-12pt}
\section*{Data availability}
{\tt DCAlign v1.0.} is available at \url{https://github.com/infernet-h2020/DCAlign}
\bibliographystyle{natbib}
\bibliography{dcalign}

\end{document}